# Configurable Distributed Physical Downlink Control Channel for 5G New Radio: Resource Bundling and Diversity Trade-off

Honglei Miao and Michael Faerber
Intel Deutschland GmbH
Am Campeon 10-12, 85579, Neubiberg, Germany
{honglei.miao, michael.faerber}@intel.com

*Abstract*—New radio technologies for the fifth generation of wireless system have been extensively studied globally. Specifically, air interface protocols for 5G radio access network will be standardized in coming years by 3GPP. Due to its crucial function in scheduled system, physical layer downlink control channel (PDCCH) is a core element to enable all physical layer data transmissions. Recently, configurable distributed PDCCH with the intention to cope with different scenarios has been developed in 3GPP. To have comprehensive understanding of respective technical advantages and potential scenario dependent limitations, detailed performance analysis and evaluations of configurable distributed PDCCH are thoroughly studied in this paper. In particular, exponential effective SNR mapping (EESM) has been employed as the performance metric of configurable distributed PDCCH in different scenarios. It is demonstrated from EESM results that configurable distributed PDCCH offers additional degree of freedom for the trade-off between achieved frequency diversity and channel estimation gain by adjusting resource bundling level according to the channel and interference scenario experienced by the control channel transmission.

I. INTRODUCTION

Currently, 5G new radio (NR) technologies are being studied in standardization body 3GPP, and selected technology as outcome of the study will be submitted to ITU as the candidate radio transmission technique for the 5th generation wireless system. Due to the successful development of vast amount of advanced techniques in current generation 3GPP standard, i.e., LTE/LTE-A, many basic physical layer aspects such as waveform and data scheduling procedures and so on adopted in LTE shall be further enhanced to meet the new requirements of 5G radio system [1]. Similar to LTE, new radio physical downlink control channel (NR-PDCCH) has been agreed in [2] to perform downlink data scheduling and uplink data assignment. This means all downlink control information (DCI) relevant to data scheduling, such as parameters used for radio resource allocation, link adaptation, hybrid ARQ and advanced MIMO operation etc, are conveyed by NR-PDCCH.

To achieve stable system operation, it is very important to ensure the reliable transmission of control channel. Motivated by frequency diversity, distributed NR-PDCCH transmission over the configured control resource set (CORESET) consisting of a number of resource blocks (RB) of 12 consecutive subcarriers or resource elements (RE) in frequency and several OFDM symbols in time, has been adopted in recent 3GPP development.

A NR-PDCCH is comprised of several resource element groups (REG), which corresponds to one RB in frequency and one OFDM symbol in time. To balance the channel estimation performance and achieved frequency diversity, REGs of distributed NR-PDCCH are further grouped into several REG bundles (REGB), each of which consists of several consecutive REGs in frequency domain using same precoding, and different REGBs are distributed over the CORESET. Given the total number of REGs in a NR-PDCCH, apparently more REGs per REGB are used, the better channel estimation performance and less frequency diversity be achieved. As such, configurable number of REGs per REGB, e.g., 2, 3 or 6, has been adopted in 3GPP. However, it is an interesting question which REGB size shall be used for a distributed NR-PDCCH in a particular scenario. To address this problem, it is required to understand the performances of these various options under different channel conditions. This paper studies the performances of different REGB constructions in channels with different frequency and interference selectivity. This would provide the insight on which REGB design option among possible alternatives shall be chosen given a particular channel condition and CORESET allocation.

The paper is organized as follows. In Section II basic control channel structure and problem statements are presented. In Section III channel estimation and performance evaluation metric are detailed. In Section IV simulation results with respect to proposed performance metric under different channel conditions are provided. Finally Section V concludes the paper.

II. CONTROL CHANNEL STRUCTURE AND PROBLEM STATEMENTS

In this section, basic distributed control channel structure and detailed problem are described.

## A. Distributed Control Channel Structure

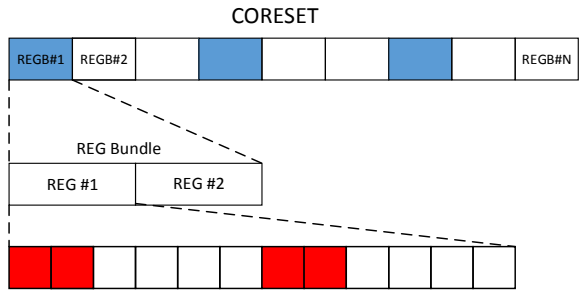

Fig. 1. CORESET of N REGBs, each REGB is comprised of 2 REGs, and each REG consists of 12 REs; REs in red are DMRS.

In 3GPP NR [2], like in LTE [3], control channel entity (CCE), consisting of 6 REGs, is defined as smallest unit of a scheduled PDCCH transmission, i.e., the granularity of the PDCCH link adaptation. As such, each NR-PDCCH is transmitted by using one or several CCEs depending on channel quality estimation of the respective UE at the base station. The number of CCEs employed by a PDCCH is called aggregation level (AL) of the PDCCH.

As shown in Fig. 1, each UE can be configured with one or several CORESETs. In case of distributed PDCCH transmission, the CORESET is comprised of a number of REGBs. Depending on configured number of REGs per REGB, i.e., 2 or 3 or 6, distributed CCE consists of various number of REGBs distributed over the RBs in the CORESET.

## B. Problem Statements

Let $N_T$ and $N_R$ define the numbers of transmit and receive antennas, respectively. We assume the channels between different transmit-receive antenna pairs are independent each other. $N_{REG}^{REGB} \in \{2,3,6\}$ and $N_{REGB}^{CCE} = 6/N_{REG}^{REGB} \in \{1,2,3\}$ denote the number of REGs/REGB and REGBs/CCE, respectively. $N_{RB}^{CORESET}$ denotes the number of resource blocks in the CORESET. Let $m, m = 1,2,...,N_{REGB}^{CCE}$ define the index of REGB in a CCE. $g_m \in C^{N_T}$ denotes the precoding vector used in the $m$th REGB of the CCE. Let $D$ denote the number of DMRS subcarriers in a REGB, and $r = (r_1,...,r_D), 1 \leq r_k \leq 12N_{REG}^{REGB}, k = 1,...,D$ defines the set of subcarrier indices of DMRS in a REGB. $x_{m,R} = (x_{m,r_1},...,x_{m,r_D})^T$ denotes the DMRS vector of the $m$th REGB of the CCE. $X_{m,R} = \text{diag}(x_{m,R})$ defines the diagonal matrix constructed from $x_{m,R}$. Moreover, $y_{m,i,R} \in C^D$ stands for the receive DMRS vector of the $m$th REGB of the CCE from the $i$th receive antenna. Let $\sigma_{z_{m,i}}^2$ denotes the noise variance at the $m$th REGB of the CCE from the $i$th receive antenna, and $z_{m,i,R} \sim CN(0, \sigma_{z_{m,i}}^2 I_D)$ refers to noise vectors accordingly. Due to the fact that interferences from neighbor cells' traffic to different REGBs can be different, the noise variance at different REGBs can be different. Let $H_{m,i} \in C^{D \times N_T}$ define the channel matrix, in which the entry $h_{m,i}^{k,n}$ at the $k$th row and $n$th column refers to the channel frequency coefficient of the $k$th DMRS of the $m$th REGB from $n$th transmit antenna to the $i$th receive antenna. Moreover, $H_{m,i}$ can be expressed as follows

$$H_{m,i} = \begin{pmatrix} (h_{m,i}^{r_1})^T \\ \vdots \\ (h_{m,i}^{r_D})^T \end{pmatrix},$$

where $h_{m,i}^{r_k} \in C^{N_T}$ stands for channel frequency coefficient vector of the $k$th DMRS of the $m$th REGB from from $N_T$ transmit antennas to the $i$th receive antenna.

Hence, the effective precoded channel vector $\bar{h}_{m,i} \in C^D$ for DMRSs in the $m$th REGB at the $i$th receive antenna can be defined as follows

$$\bar{h}_{m,i} = H_{m,i} g_m \quad (1)$$

Then we have the following expression for the received DMRS signal vector of the $m$th REGB of the CCE from the $i$th receive antenna

$$y_{m,i,R} = X_{m,R} \bar{h}_{m,i} + z_{m,i,R} \quad (2)$$

Given the signal models in Eqs. (1) and (2), specifically, the size of CORESET, i.e, $N_{RB}^{CORESET}$, which determines control channel capacity and the locations of REGBs of each distributed CCE in CORESET, an interesting problem is to find the optimal value of $N_{REG}^{REGB}$, among 2, 3 and 6 for a given channel power delay profile (PDP), e.g., TDL-A/B/C in [4] and interference scenario to achieve best receive performance of CCE in terms of block error rate (BLER). It is known that BLER performance of distributed CCE depend on multiple facets including applied channel coding scheme, channel estimation errors depending on the REGB size $N_{REG}^{REGB}$ used for MMSE filtering, and frequency diversity obtained by $N_{REGB}^{CCE}$ which is also determined by $N_{REG}^{REGB}$.

Given instantaneous SNRs at different subcarriers/resource blocks, exponential effective SNR mapping (EESM) has been widely used [5] as an important performance metric to predict the link BLER performance from multiple channels with different SNRs. With the signal model in Eq. (2), we study the EESMs of distributed CCE with various values of $N_{REG}^{REGB}$ for different channel and interference scearios. To account for the channel estimation errors affected by $N_{REG}^{REGB}$, we employ estimated EESM based on MMSE channel estimates obtained from Eq. (2).

## III. CHANNEL AND EESM ESTIMATION

In this section, MMSE based channel estimation and EESM calculation are presented.

### A. MMSE Based Channel Estimation

Let $p = (p_1,...,p_L)$ define the channel power delay profile, and $\tau = (\tau_1,...,\tau_L)$ define the channel tap delays in the resolution of sample duration of OFDM symbols. Let $\tau_S$ denote the sample duration. Moreover, $h_k, k = 0,...,K-1$, where $K$ stands for the total number of subcarriers in an OFDM symbol, defines channel frequency coefficients vector of the $k$th subcarrier of the channel. With simple

mathematical derivation, we can obtain the following expression

$$R_h(\Delta k) = E(h_k h_{k-\Delta k}^*) = \sum_l p_l e^{-i\frac{2\pi(\Delta k)\tau_l}{K\tau_S}}$$

Due to the assumption that receive channels at different receive antennas are independent, channel estimation for each receive antenna can be performed independently. From Eq. (2), it is well known that the MMSE based channel estimates $\hat{\bar{h}}_{m,i}$ can be obtained by the following expression

$$\begin{aligned}\hat{\bar{h}}_{m,i} &= R_{\bar{h}_{m,i}, y_{m,i,R}}\left(R_{y_{m,i,R}}\right)^{-1} y_{m,i,R} \\ &= R_{\bar{h}_{m,i}} X_{m,R}^H \left(X_{m,R} R_{\bar{h}_{m,i}} X_{m,R}^H + \sigma_{z_{m,i}}^2 I_L\right)^{-1} y_{m,i,R}\end{aligned} \quad (3)$$

where the $(u,v)$th entry of $R_{\bar{h}_{m,i}}$ can be further calculated as follows

$$\begin{aligned}R_{\bar{h}_{m,i}}^{(u,v)} &= E\left(\left(h_{m,i}^{r_u}\right)^T g_m \left(\left(h_{m,i}^{r_v}\right)^T g_m\right)^H\right), \\ &= g_m^H E\left(\left(h_{m,i}^{r_u}\right)^* \left(h_{m,i}^{r_v}\right)^T\right) g_m, \\ &= g_m^H \left(E\left(\left(h_{m,i}^{r_u}\right)\left(h_{m,i}^{r_v}\right)^H\right)\right)^* g_m, \\ &= g_m^H \left(E\left(\left(h_{m,i}^{r_u,n}\right)\left(h_{m,i}^{r_v,n}\right)^H\right)\right)^* I_{N_T} g_m, \\ &= R_h(r_u - r_v).\end{aligned}$$

*B. EESM Calculation*

Let $\gamma_{m,i,j} = \frac{E(|\bar{h}_{m,i,j}|^2)}{\sigma_{z_{m,i}}^2}$ define the SNR of the $j$th received DMRS subcarrier of the $m$th REGB of the NR-PDCCH at the $i$th receive antenna, received DMRS from all receive antennas can be further combined according to respective SNR of each receive antenna. As a result, $\gamma_{m,j} = \sum_{i=1}^{N_R} \gamma_{m,i,j}$ denote the combined SNR of the $j$th received DMRS subcarrier of the $m$th REGB of the NR-PDCCH. With the definition of EESM from [5], EESM based on SNRs experienced at $N_{RS} = D N_{REGB}^{CCE}$ DMRS subcarriers of the NR-PDCCH can be expressed as

$$\gamma_{\text{eff}} = -\lambda \ln\left(\frac{1}{N_{RS}} \sum_{m=1}^{N_{REGB}^{CCE}} \left(\sum_{j=1}^{D} \exp\left(-\frac{\gamma_{m,j}}{\lambda}\right)\right)\right), \quad (4)$$

where $\lambda$ is an adjustment scaler [5] selected experimentally for different code rates of a particular coding scheme. It is clear from Eq. (3) that MMSE channel estimate $\hat{\bar{h}}_{m,i}$ requires the knowledge about $R_{\bar{h}_{m,i}}$ and $\sigma_{z_{m,i}}^2$. Therefore, prior to MMSE channel estimation, estimation of $R_{\bar{h}_{m,i}}$ and $\sigma_{z_{m,i}}^2$ needs to be performed. For example, least-squared channel estimates can be used for power delay profile and noise variance estimation. With the enhanced MMSE channel estimates from Eq. (3), SNR estimate $\hat{\gamma}_{m,i,j}$ can be further improved as follows

$$\hat{\gamma}_{m,i,j} = \frac{E\left(\left|\hat{\bar{h}}_{m,i,j}\right|^2\right)}{E\left(\left|y_{m,i,j,R} - x_{m,j,R}\hat{\bar{h}}_{m,i,j}\right|^2\right)} \quad (5)$$

## IV. SIMULATION RESULTS AND DISCUSSIONS

In this section, computer simulations have been conducted to study the EESM performance of different REGB design options, i.e., $N_{REG}^{REGB} = \{2,3,6\}$, which can achieve various trade-off between diversity and bundling gain (channel estimation gains) for distributed NR-PDCCH. Two interference scenarios, i.e. frequency-flat and frequency-selective, are simulated. In frequency-flat interference scenario, all RBs in CORESET experience the same SINR. However in frequency-selective interference scenario, some RBs in CORESET suffer from stronger interference than other RBs. Such frequency-selective interference scenario imitates the situation where strong interferences are caused by localized data or control transmission in neighbor cells.

The benefits from frequency diversity transmission can be easily manifested in frequency-selective interference scenario. Moreover, to demonstrate different interference selectivity, two different amount of RBs suffering from stronger interferences, i.e., one quarter and one half of CORESET, are further investigated. The detailed simulation parameters are listed in Table 1.

TABLE I. SIMULATION PARAMETERS

| Parameters | Values |
|---|---|
| Channel model | TDL-A [4], delay scaling factor: 300ns |
| Number of base station transmit antennas | 2 |
| Number of UE receive antennas | 1 |
| Subcarrier spacing [kHz] | 15 |
| Symtem bandwidth [MHz] | 20 |
| Control resource RB set | [1:48] |
| Number of control symbols | 1 |
| RBs with 3dB stronger interferenes | [1:12] and [1:24] |
| Number of REGs/CCE | 6 |
| Number of REGs/REGB | 2, 3 and 6 |
| Aggregation levels | 1, 2, 4 and 8 |
| Transmission scheme | Per-REGB random precoder cycling. |
| Channel estimation | MMSE, averaging over REGB |

*A. Stronger Interference in First Quarter of CORESET*

In frequency-flat interference scenario, the interference powers at all RBs of CORESET are given by the x-axis in Figs. 2-5, and referred by the legend with "$k$REGs/REGB, 0dB" , where $k$=2,3,6, denoting the number of REGs per REGB. In frequency-selective interference scenario, the interference powers in the first 12 RBs of CORESET are 3 dB stronger than that for other RBs, and referred by the legend with "$k$REGs/REGB, 3dB".

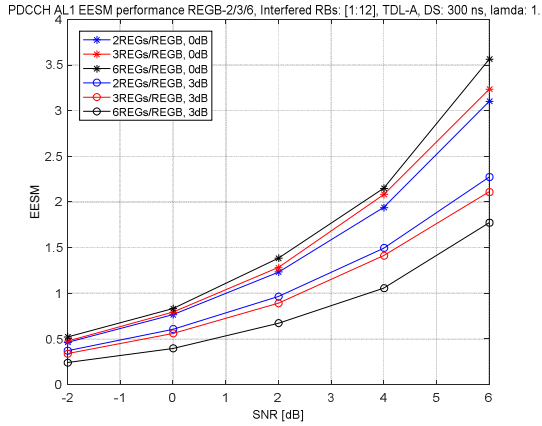

Fig. 2. PDCCH AL1, RBs with 3dB stronger interferences: [1:12], TDL-A, DS: 300ns.

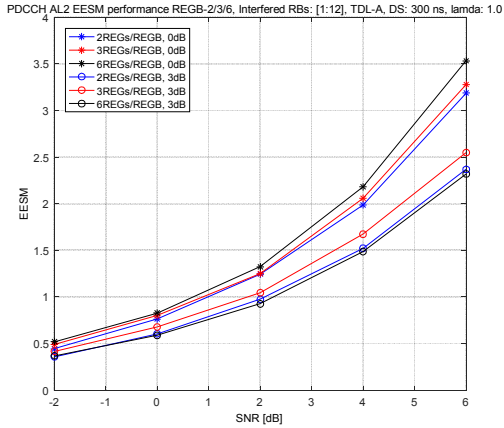

Fig. 3. PDCCH AL2, RBs with 3dB stronger interferences: [1;12], TDL-A, DS: 300ns.

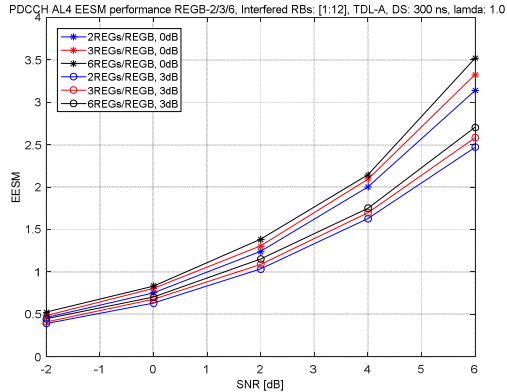

Fig. 4. PDCCH AL4, RBs with 3dB stronger interferences: [1:12], TDL-A, DS: 300ns.

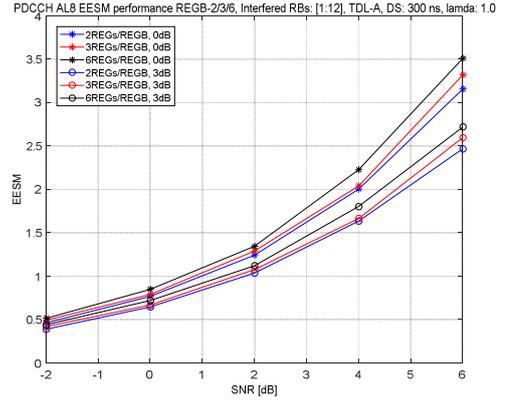

Fig. 5. PDCCH AL8, RBs with 3dB stronger interferences: [1:12], TDL-A, DS: 300ns.

The EESM results of PDCCHs with AL1, 2, 4 and 8 are illustrated in Figs. 2-5, respectively. In addition to more coding gain, due to the distributed transmission, PDCCHs with high ALs benefit more frequency diversity as well. It is shown from Figs. 2-5 that due to less interferences, EESM of PDCCH in frequncy-flat interference scenario outperforms those in frequncy-selective interference scenarios. In frequency-flat interference scenario, it seems that no additional frequency diversity can be obtained by smaller REGB size, however, due to better bundling/channel estimation gain, the REGB of 6REGs exhibits the best performance among three REGB designs.

It is from Fig. 2 that in frequency-selective interference scenario, for AL1 PDCCH, the amount of REs suffeing from stronger interferences increases according to the number of REGs per REGB of the PDCCH. In other words, more diversities are obtained by 2 and 3 REGs/REGB based PDCCH. Hence, REGB of 2REGs shows the best EESM performance. However it is observed from Fig. 3 that for AL2 PDCCHs, despite of more REs of PDCCH based on 3 REGs/REGB suffering from higher interferences than that based on 2 REGs/REGB, the bundling gain of 3 REGs/REGB offsets the performance loss caused by the less frequency diversity. As a result, the 3 REGs/REGB exhibits the best performance in frequency-selective interference scenario. Due to the similar reason, it is shown from Figs. 4-5, the bundling gain from 6 REGs/REGB are more prominent than the loss due to the less frequency diversity, high AL PDCCH with 6 REGs/REGB demonstrates the best performance among different REGB designs.

*B. Stronger Interference in First Half of CORESET*

To invesigate the impact of different interference selectivity on EESM performance, we simulate a different subset of RBs, i.e., [1:24], in the CORESET suffering from stronger interferences. As such, the resulted high interference RBs constitute of half of the CORESET. The EESM performances of AL1 and 2 PDCCHs are illustrated in Figs. 6 and 7, respectively.

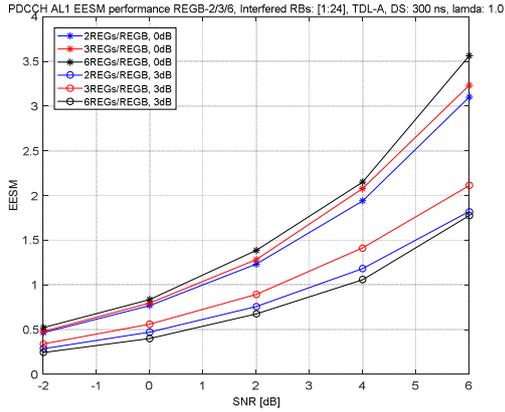

Fig. 6. PDCCH AL1, RBs with 3dB stronger interferences: [1:24], TDL-A, DS: 300ns.

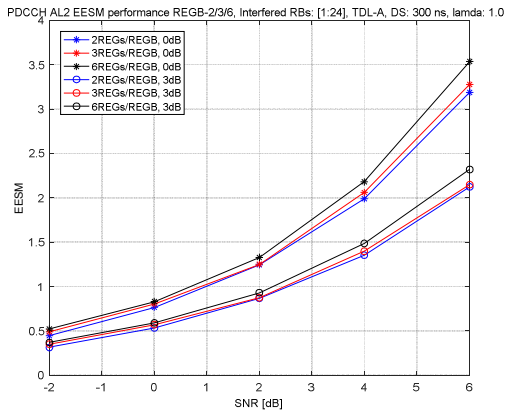

Fig. 7. PDCCH AL2, RBs with 3dB stronger interferences: [1:24], TDL-A, S: 300ns.

For AL1 PDCCH, it is shown from Fig. 6 that due to a larger subset of RBs in CORESET suffering from stronger interference, the amount of victim REs in 3 REGs/REGB are even less that that of 2 REGs/REGB based PDCCH, the PDCCH with 3 REGs/REGB shows the best performance among three REGB designs. However for AL2 PDCCH, it is shows from Fig. 7 that the strong bundling gain from 6 REGs/REGB offers more benefits than the loss due to the smaller frequency diversity. As a result, the PDCCH based on 6 REGs/REGB achieves the best EESM performance.

*C. Discussions*

Based on the above simulation results, it shows that for frequnecy-flat interference scenarios, where whole CORESET experiences same interfernce level, the channel estimation gain from REG bundling offers the most prominent benefits with respect to EESM performance, as such 6 REGs/REGB seems to be the most favorable REGB design option. Such effects remain for high AL PDCCHs in frequency-seletive interference scenarios. For low AL PDCCH, e.g., AL1, and highly frequency-selective interference scenario, 2 and 3 REGs/REGB seem to offer better EESM performance. Overall 3 REGs/REGB seem to be more robust solution than 2 REGs/REGB.

## V. CONCLUSIONS

This paper studies the EESM performance of configurable distributed NR-PDCCH with different REG bundle sizes. Different REG bundle sizes offer different trade-off between channel estimation performance gain and achieved frequency diversity. Specifically, different interference scenarios, i.e., frequency-flat interference and frequency-selective interferences are investigated. It is demonstrated from simulation results that large REG bundle size provides best overall EESM performance in frequency-flat interference scenario where channel estimation performance plays the dominant role on the reception performance of NR-PDCCH. However, in frequency-selective interference scenarios, where diversity transmission is more beneficial for the reliable reception of PDCCH. As a consequence, small and mediate REG bundle sizes furnish better EESM performance and are more preferable. This clearly motivates the configurability of REG bundle size for NR-PDCCH operating in different interference scenarios.

ACKNOWLEDGMENT

The research leading to these results received funding from the European Commission H2020 programme under grant agreement n°760809 (ONE5G project).